\documentclass[twocolumn,showpacs,preprintnumbers,superscriptaddress,prb,aps]{revtex4-1}

\usepackage{amsmath,amssymb}
\usepackage{txfonts}
\usepackage{textcomp}
\usepackage[pdftex]{graphicx}
\graphicspath{{fig/}}

\begin{document}
\title{Ferroic quadrupolar ordering in CeCoSi
    revealed using $^{59}$Co-NMR measurements}

\author{Masahiro Manago}
\email{manago@riko.shimane-u.ac.jp}
\affiliation{Department of Applied Physics, Shimane University,
    Matsue 690-8504, Japan}
\author{Ayano Ishigaki}
\author{Hideki Tou}
\author{Hisatomo Harima}
\affiliation{Department of Physics, Kobe University, Kobe 657-8501, Japan}
\author{Hiroshi Tanida}
\affiliation{Liberal Arts and Sciences, Toyama Prefectural University,
    Imizu, Toyama 939-0398, Japan}
\author{Hisashi Kotegawa}
\affiliation{Department of Physics, Kobe University, Kobe 657-8501, Japan}

\begin{abstract}
    A nonmagnetic phase transition at $T_0 \sim 12$ K in the tetragonal system
    CeCoSi with a Kramers doublet ground state is reminiscent of an electric
    quadrupole ordering, even though its well-separated crystal-electric-field
    (CEF) levels are unlikely to acquire higher-order multipole degrees
    of freedom.
    Here, we report $^{59}$Co nuclear magnetic resonance (NMR) studies that are
    highly compatible with a ferroic quadrupole ordering below $T_0$.
    Changes in the NMR spectra below $T_0$ suggest that an external magnetic
    field induces ferroic Ce dipole moments orthogonal to the field,
    enabling domain selection in the nonmagnetic phase.
    Our findings suggest the presence of a ferroic $O_{zx}$-type quadrupole
    component in CeCoSi and demonstrate that quadrupole ordering
    may occur under well-separated CEF levels in tetragonal systems.
\end{abstract}

\maketitle

\section{Introduction}

Order parameters in phase transitions are determined through crystal symmetries
and interactions.
In $f$-electron systems, the degrees of freedom of unpaired electrons are
strongly affected by spin--orbit coupling,
but ultimately, the crystal electric field (CEF) determines their ground state.
If a CEF ground state is a Kramers doublet,
the $f$-electron degrees of freedom are limited to a magnetic  origin unless the
excited states are located closely
\cite{JPSJ.68.2057,PhysRevB.73.020407,PhysRevB.48.256,PhysRevB.58.6339,PhysRevLett.122.187201,NaturePhys.16.546}.
This restriction is modified by the hybridization between the conduction
electrons and the $f$-electrons ($c$--$f$ hybridization);
Kondo lattice systems that have the Kramers doublet as their ground state have
a chance to acquire higher-order multipole degrees of freedom owing to the
contribution of the CEF excited states.
A rare example of this is CeTe \cite{JPSJ.80.023713}.
Although CeTe is a cubic system, the CEF ground state is a Kramers doublet,
and the separation between the ground state and the excited state is
$\Delta \sim 30$ K.
Under pressure, the enhanced Kondo effect induces an electric quadrupole
ordering, even though $\Delta$ is one order larger than the quadrupole
ordering temperature.
Another possible example has been recently demonstrated in the tetragonal system
CeRh$_2$As$_2$ with the Kramers doublet ground state
\cite{science.abe7518,PhysRevX.12.011023}.
A quadrupole-density-wave state below 0.4 K has been proposed, although
$\Delta \sim 30$ K is two orders larger than the ordering temperature.
In CeRh$_2$As$_2$ and CeTe, the Kondo temperatures $T_{\textrm{K}}$ are
comparable to $\Delta$, indicating the sufficient mixing of two CEF levels.

Here, we focus on the tetragonal system CeCoSi, in which quadrupole ordering has
been proposed \cite{JPSJ.87.023705,JPSJ.88.054716}.
In CeCoSi, aside from the antiferromagnetic transition at Néel temperature
$T_{\textrm{N}} = 9.4$ K, another phase transition was initially observed under
pressure; the transition temperature of $T_0 = 38$ K at 1.5 GPa
\cite{PhysRevB.88.155137}.
This was later also confirmed at ambient pressure at $T_0 = 12$ K
\cite{JPSJ.88.054716}.
The increase in $T_0$ under the magnetic field was reminiscent of the quadrupole
ordering \cite{JPSJ.88.054716}, whereas $\Delta$ is $\sim 125$ K
\cite{PhysRevB.101.214426}.
This value is much higher than those reported for CeTe and CeRh$_2$As$_2$
and comparable to that of the prototype Ce-based tetragonal system CeCu$_2$Si$_2$
($\Delta \sim 140$ K) \cite{PhysRevB.23.3171}, where quadrupole degrees of
freedom have been inconsiderable.
Although interorbital interactions between the ground and excited states can
induce quadrupole orderings phenomenologically even with such a large
$\Delta$ \cite{JPSJ.89.013703}, their feasibility remains an open question.

Our previous $^{59}$Co-nuclear magnetic resonance (NMR) and
nuclear quadrupole resonance (NQR) measurements in CeCoSi show the symmetry
breaking of a nonmagnetic origin below $T_0$ \cite{JPSJ.90.023702}.
The splitting of the NMR spectrum suggested the emergence of the
field-induced dipole moments, which are characteristic in quadrupole ordering
states \cite{JPSJ.52.728,JPSJ.66.1741,PhysRevLett.94.137209,JPSJ.85.113703}.
However, the combined results of NMR and NQR did not reveal the order
parameter.
A recent high-resolution x-ray diffraction (XRD) measurement has clearly
revealed the triclinic distortion below $T_{\textrm{s1}}$, which is
considered the same as $T_0$ at zero field \cite{JPSJ.91.064704}.
The triclinic angles are $\alpha=\beta=89.64$\textdegree\ and
$\gamma\simeq90$\textdegree\ at 10 K.
If we recognize it as the quadrupole ordering, this distortion is compatible
with a ferroic $O_{yz}+O_{zx}$-type ordering.
The interpretation of the previous NMR study was inconsistent with this
ferroic ordering because the splitting of the spectrum was reminiscent
of antiferroic ordering.
Solving this inconsistency is essential to understand the nature of the
nonmagnetic ordered phase in CeCoSi.

In this article, we present field-angle-controlled NMR spectra measured
at ambient pressure on a CeCoSi single crystal.
In the nonmagnetic ordered state, we found that one of the split NMR peaks
became predominant as tilting the external magnetic field from the in-plane
direction.
This suggests that the splitting arises not from the emergence of inequivalent
Co sites in a unit cell but from different domains of a ferroic ordered state.
The results of this study eliminate the inconsistency between XRD and the previous
NMR findings, being very compatible with the presence of a ferroic $O_{zx}$
quadrupole component below $T_0$.

\section{Experimental}

A plate-shaped single-crystalline CeCoSi sample (4 mm $\times$ 2 mm $\times$
0.4 mm) was grown using the Ce/Co eutectic flux method, as described in
Ref.~\onlinecite{JPSJ.88.054716}.
$^{59}$Co (nuclear spin $I=7/2$) NMR measurements were performed at ambient
pressure and a temperature range of 10--20 K above $T_{\textrm{N}}$.
NMR spectra were measured in several field directions controlled by a home made
double-axis rotator.
The field angle was evaluated from the NMR frequencies above $T_0$.
The angle precision is $\sim 0.1$\textdegree\ for the polar angle $\theta$ and
$\sim 5$\textdegree\ for the azimuth angle $\phi$.
The electric field gradient (EFG) parameters, i.e, the quadrupole frequency
$\nu_{\text{Q}}$ and the asymmetry parameter $\eta$, were deduced from the NQR results \cite{JPSJ.90.023702}
(see also the Appendix) and fixed in the analysis of the NMR spectra.
The EFG at the Co site was calculated through a
full-potential linear augmented plane wave (LAPW) calculation within the local
density approximation (LDA).

\section{Results and discussion}
\begin{figure}
    \centering
    \includegraphics[width=85mm]{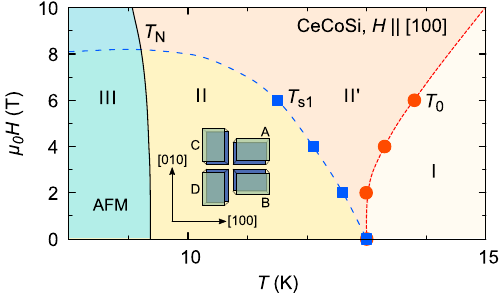}
    \caption{\label{fig:phasediagram}
        Field--temperature phase diagram of CeCoSi along the $[100]$ direction.
        The transition temperatures $T_0$ and $T_{\textrm{s1}}$ and the N\'eel
        temperature $T_{\textrm{N}}$ were from literature
        \cite{JPSJ.91.064704,JPSJ.88.054716}.
        Phases II and II' represent the nonmagnetic ordered states.
        The crystal symmetry is lowered from tetragonal (phase I) to triclinic
        (phase II), and that in phase II' remains unclear.
        The inset shows four domains for the triclinic distortion below
        $T_{\textrm{s1}}$\cite{JPSJ.91.064704}.
    }
\end{figure}

First, we show the field--temperature phase diagram in
Fig.~\ref{fig:phasediagram}, which was revealed by XRD \cite{JPSJ.91.064704}
and confirmed by bulk measurements \cite{JPSJ.91.094701}.
Here, $T_0$ increased by applying the external field along the [100] direction,
while the structural transition temperature $T_{\textrm{s1}}$ decreased.
The separation of $T_0$ and $T_{\textrm{s1}}$ suggests multiple components of
the order parameter.
The inset shows the four domains (A--D) in the triclinic symmetry confirmed in
phase II \cite{JPSJ.91.064704}, while a change in the structural symmetry in
phase II' is undetected at the moment.

\begin{figure}
    \centering
    \includegraphics{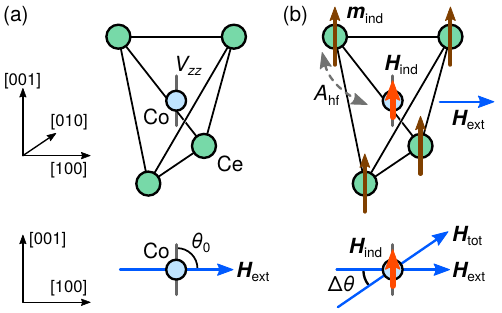}
    \caption{\label{fig:crystal-inducedfield}
    The arrangement of the NMR measurement for CeCoSi.
    (a) A Co site surrounded by four Ce atoms forming a tetrahedron.
    The maximum principal axis of the EFG, $V_{zz}$, is parallel to the $[001]$
    direction.
    The bottom figure shows the magnetic field direction $\theta_0$ from
    $[001]$.
    (b) Below $T_0$, induced moments at the Ce sites $\bm{m}_{\textrm{ind}}$
    emerge perpendicular to the external field.
    The magnetic field at the Co site $H_{\textrm{ind}}$ is induced through
    the hyperfine coupling.
    Only the perpendicular component of the Ce moment is shown.
    The bottom figure shows the change in the field angle $\Delta\theta$ due to
    the emergence of $H_{\textrm{ind}}$.
    }
\end{figure}

\begin{figure}
    \centering
    \includegraphics{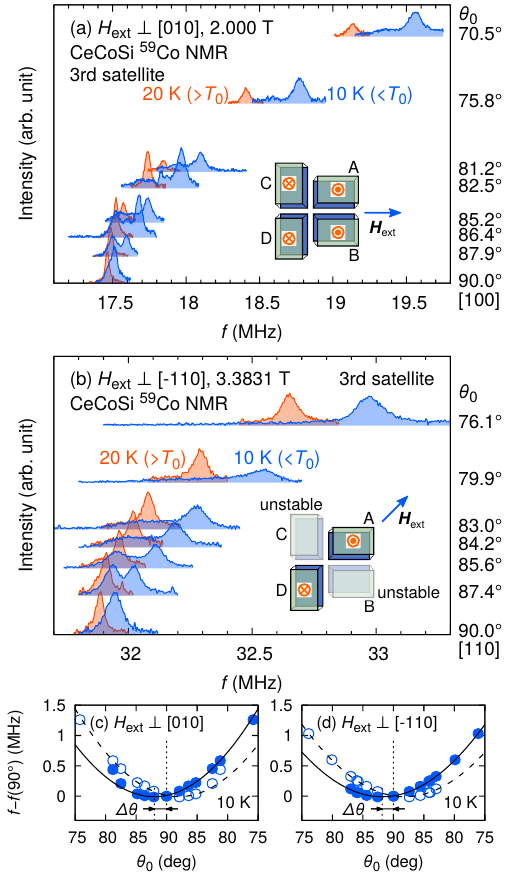}
    \caption{\label{fig:angle-fsp}
        (a) Field-angle dependence of the $^{59}$Co NMR third satellite spectra
        of CeCoSi at 10 and 20 K with $\theta_0$ from $[001]$ to the $[100]$
        axis at 2.000 T.
        The inset represents the induced field directions shown in red for
        each domain.
        (b) Field-angle dependence between $[110]$ and $[001]$ directions
        at 3.3831 T.
        Some domains are unstable for the in-plane field.
        In both cases, the domain selection occurs in the out-of-plane field.
        (c), (d) Field angle dependence of the resonant frequencies at 10 K
        for (c) $H \perp [010]$ and (d) $H \perp [\bar{1}10]$.
        The solid and dashed lines show the best fit of the result with
        $\frac{9}{2}\nu_{\textrm{Q}}\cos^2(\theta_0 \pm \Delta\theta)$.
    }
\end{figure}

The Co site in CeCoSi is surrounded by the four nearest Ce atoms, forming a
tetrahedron, as shown in Fig.~\ref{fig:crystal-inducedfield}(a).
The local symmetry is $\bar{4}m2$ above $T_0$, ensuring a single site in the
NMR spectrum for any field directions.
In this symmetry, the maximum principal axis of the EFG, $V_{zz}$, is directed
along the [001] direction at the Co site.
The field-angle dependence of the NMR spectra at the third satellite peak
($7/2 \leftrightarrow 5/2$ transition) is shown in Fig. \ref{fig:angle-fsp}(a).
The measurements at 10 K were performed in phase II, where the four domains
were formed.
The direction of the magnetic field varied from $[100]$ to $[001]$.
Here, $\theta_0$ is the angle between the external magnetic field,
$H_{\textrm{ext}}$, and the [001] direction in the tetragonal symmetry.
At 10 K, when $\theta_0$ was slightly changed from 90\textdegree,
the peak splitting occurred due to the symmetry lowering below $T_0$
\cite{JPSJ.90.023702}.
With further tilting, the intensity of the lower-frequency peak became weaker
and almost disappeared at $\theta_0 \lesssim 80$\textdegree.
This strongly suggests that the NMR splitting arises not from the emergence of
inequivalent Co sites in a unit cell but from a difference in the resonance
condition among the domains; the Co site remained equivalent in each domain,
and some of the A--D domains were selected by the $c$-axis component of
the magnetic field.
Similar results were obtained between the [110] and [001] directions,
as shown in Fig.~\ref{fig:angle-fsp}(b).
Two peaks were observed when the magnetic field was tilted slightly
from the [110] direction.
These correspond to the domains A and D because the XRD study clarified the
domain selection by the in-plane field along the [110] direction
\cite{JPSJ.91.064704}.
The NMR results revealed further domain selection induced by the out-of-plane
field at $\theta_0 \lesssim 80$\textdegree.

Figures \ref{fig:angle-fsp}(c) and \ref{fig:angle-fsp}(d) show the field-angle
dependence of the peak frequencies at 10 K.
The results are symmetrized with respect to $\theta_0=90$\textdegree.
They are reproduced by
\begin{equation}
    f = \gamma H_{\textrm{ext}} [1 + K(\theta_0)]
    + \frac{3}{2} \nu_{\textrm{Q}} (3 \cos^2 \theta - 1),
\end{equation}
which is obtained within the first-order perturbation with respect to the
quadrupolar frequency $\nu_{\textrm{Q}}$,
where $\gamma$ is a gyromagnetic ratio, $K(\theta_0)$ is the Knight shift, and
$\theta$ is the angle between the total magnetic field, $H_{\textrm{total}}$,
at the Co site and the direction of $V_{zz}$.
Above $T_0$, $\theta$ is equal to $\theta_0$ determined by $H_{\textrm{ext}}$,
and the frequency achieves its minimum at $\theta_0 = \theta = 90$\textdegree.
The symmetry reduction below $T_0$ causes a difference between $\theta_0$ and
$\theta$, which is denoted as $\theta = \theta_0 \pm \Delta\theta$.
The best fit was obtained with a tilting angle of
$\Delta\theta \simeq 2$--3\textdegree.
This means that the NMR splitting is due to the change in the angle between
$H_{\textrm{total}}$ and the $V_{zz}$ \cite{JPSJ.90.023702}.

We carefully checked the possibility that $\Delta \theta$ mainly arises
from the tilting of $V_{zz}$ due to the triclinic distortion.
Thus, the changes in the EFG parameters were obtained through the band-structure
calculation using structural parameters in triclinic symmetry
\cite{JPSJ.91.064704}.
The calculation provided the result that the tilting of $V_{zz}$ is
0.16\textdegree, one order smaller than the experimental $\Delta \theta$.
This suggests that the change in the EFG is very small in CeCoSi,
and that nonzero $\Delta\theta$ arises from the tilting of $H_{\textrm{total}}$
at the Co site.
We investigated the breaking of the four-fold symmetry in the EFG using NQR
measurements, but its change into the triclinic symmetry was undetected within
the experimental error, as shown in the Appendix.
This is also consistent with the calculation showing a small change in the EFG.

\begin{figure}[b]
    \centering
    \includegraphics[width=85mm]{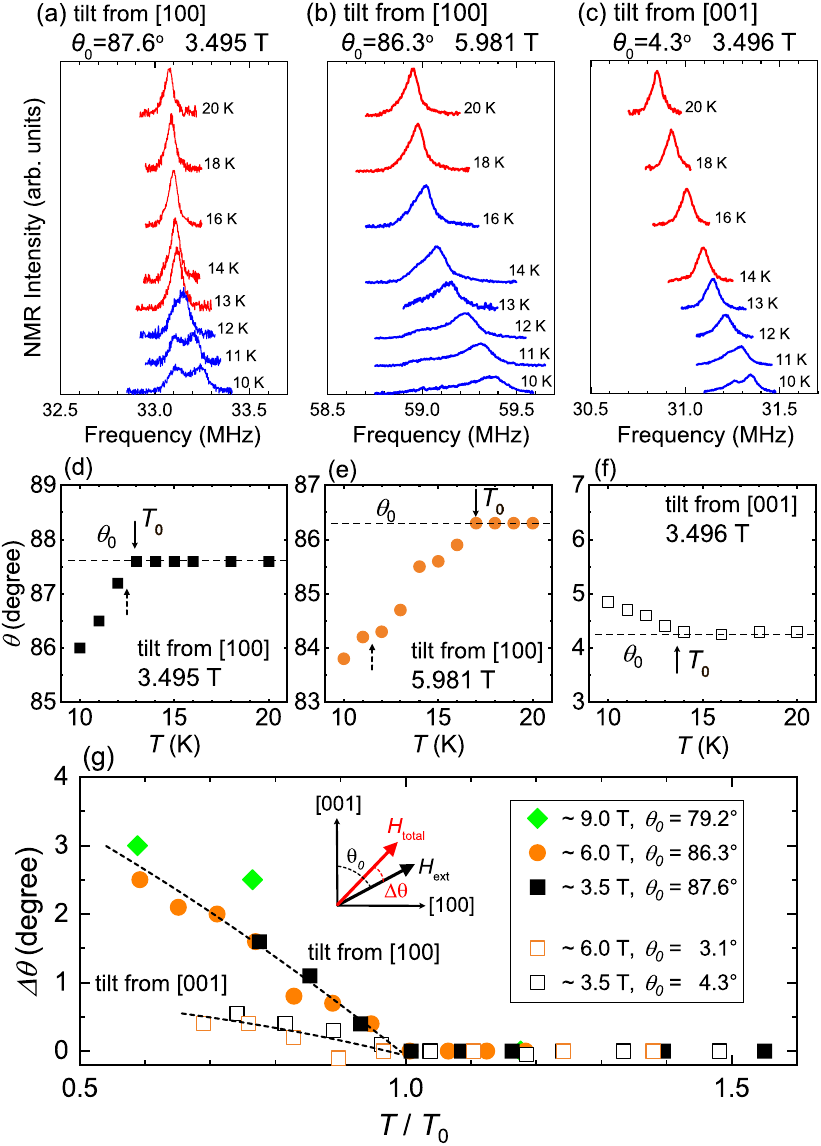}
    \caption{\label{fig:spectra}
        (a)--(c) Temperature dependence of the $^{59}$Co NMR third satellite
        peaks:
        (a) 3.495 T with $\theta_0 = 87.6$\textdegree,
        (b) 5.981 T with $\theta_0 = 86.3$\textdegree\,
        and (c) 3.496 T with $\theta_0 = 4.3$\textdegree.
        The spectra are vertically shifted for clarity.
        (d)--(f) Temperature dependence of the angle $\theta$ of $^{59}$Co NMR
        spectra in several fields.
        The horizontal dashed lines represent the external field angle
        $\theta_0$.
        The solid arrows indicate $T_0$ determined in NMR, while the dotted
        arrows are $T_{\textrm{s1}}$ \cite{JPSJ.91.064704}.
        (g) Temperature dependence of the tilting angle $\Delta\theta$ of
        $^{59}$Co NMR spectra in several fields.
        The temperature is normalized by $T_0$ for each field.
        The dotted curves are guides for the eye.
    }
\end{figure}

This interpretation is also supported by the field-direction dependence
of $\Delta\theta$ shown below.
Figures \ref{fig:spectra}(a)--\ref{fig:spectra}(c) show the temperature
dependence of the NMR spectra in different fields.
For the respective spectra, $\theta$ was evaluated for the high-frequency
peak of the split peaks.
The temperature dependence is shown in
Figs.~\ref{fig:spectra}(d)--\ref{fig:spectra}(f).
A clear kink corresponding to $T_0$ appears for all the field settings.
$T_0$ increases as $H_{\textrm{ext}}$ increases along the [100] direction,
which is in good agreement with other experimental reports
\cite{JPSJ.88.054716,JPSJ.91.064704,JPSJ.91.094701}.
$T_0$ in NMR is slightly higher than those in other reports, which is probably
owing to the out-of-plane component of the $H_{\textrm{ext}}$.
The temperature dependences of $\Delta\theta = \lvert \theta_0 - \theta \rvert$
are summarized in Fig.~\ref{fig:spectra}(g),
where the temperatures are normalized by $T_0$.
The $\Delta\theta$ is almost independent of the strength of $H_{\textrm{ext}}$
for each field direction.
At $T/T_0 \simeq 0.7$, $\Delta\theta$ for $H_{\textrm{ext}}$ tilted from [100]
is approximately four times larger than that for $H_{\textrm{ext}}$ tilted
from [001].
The origin of $\Delta\theta$ cannot be explained solely by the tilting of the
$V_{zz}$ originating in the structural distortion because it must provide the
same $\Delta\theta$ in any field directions, clearly indicating that
$\Delta\theta$ is dominated by the tilting of $H_{\textrm{total}}$.

As shown in Fig.~\ref{fig:crystal-inducedfield}(b), the tilting of
$H_{\textrm{total}}$ arises from the emergence of the $H_{\textrm{ind}}$
at the Co site, which is orthogonal to the ${H_{\textrm{ext}}}$.
Because ${H_{\textrm{ind}}} /{H_{\textrm{ext}}} \simeq \tan(\Delta\theta)$,
the field-independent $\Delta\theta$ indicates
that $H_{\textrm{ind}}$ is almost proportional to $H_{\textrm{ext}}$.
The observations of the peaks assigned to the single Co site in NMR
at $\theta_0 \lesssim 80$\textdegree\ suggest that the $H_{\textrm{ind}}$
is uniform at the Co nuclei in a domain; therefore, the Ce local moments
should be uniform below $T_0$.
In this case, the hyperfine field at the Co site is dominated by a diagonal
part \cite{JPSJ.90.023702} (see also the Appendix); that is, the $H_{\textrm{ind}}$
at the Co site is induced by the Ce moments parallel to them,
as shown in Fig.~\ref{fig:crystal-inducedfield}(b).
In the case of $H_{\textrm{ext}} \parallel [100]$, the orthogonal
$H_{\textrm{ind}}$ along the [001] direction is considered to arise from
$A_{cc}m_{\textrm{ind}}^{c}$, where $A_{ij}$ is a component of the hyperfine
coupling tensor, and $m_{\textrm{ind}}^{i}$ is the $i$-axis component of the
induced Ce moment ($i,\,j = a,\,b,$ and $c$).
Likewise, in the case of $H_{\textrm{ext}} \parallel [001]$, the orthogonal
$H_{\textrm{ind}}$ is considered to arise from $A_{aa}m_{\textrm{ind}}^{a}$.
Therefore, the $H_{\textrm{ext}}$ induces the orthogonal Ce dipole moments.
Experimentally, $\Delta\theta$ for the [100] direction is four times larger
than that for the [001], as shown in Fig.~\ref{fig:spectra}(g).
It arises from a difference between $A_{cc}m_{\textrm{ind}}^{c}$
and $A_{aa}m_{\textrm{ind}}^{a}$.
Using $A_{cc} \sim 3$ T$/\mu_{\textrm{B}}$ and
$A_{aa} \sim 0.6$ T$/\mu_{\textrm{B}}$ \cite{JPSJ.90.023702},
anisotropy in the induced moments $m_{\textrm{ind}}^{c}$ and
$m_{\textrm{ind}}^{a}$ is considered weak.
It is possible to make a rough estimation of the
orthogonal induced moment size
using the relations of $H_{\textrm{ind}} = A_{ii}m_{\textrm{ind}}^{i}$ and
${H_{\textrm{ind}}} /{H_{\textrm{ext}}} \simeq \tan(\Delta\theta)$.
It yields
\begin{equation}
    \frac{m_{\textrm{ind}}^{c}}{H_{\textrm{ext}}} \sim \frac{m_{\textrm{ind}}^{a}}{H_{\textrm{ext}}}
    \sim \frac{\tan(\Delta\theta)}{A_{ii}} \sim 0.01 \mu_{\textrm{B}}/\textrm{T}
\end{equation}
for $H_{\textrm{ext}} \parallel [100]$ and $[001]$, respectively,
when the tilting angle is $\Delta\theta \simeq 2$\textdegree\ for $H_{\textrm{ext}} \parallel [100]$
and $\Delta\theta \simeq 0.5$\textdegree\ for $H_{\textrm{ext}} \parallel [001]$.
The Ce magnetic moment also possesses the component parallel to the external field,
which is estimated from the Knight shift as
\begin{equation}
    \frac{m_{i}}{H_{\textrm{ext}}} = \frac{K_i - K_i^{\text{orb}}}{A_{ii}},
\end{equation}
where $K_i^{\text{orb}}$ is an orbital shift along the $i$ direction.
The moment is $0.025 $ and $0.030 \mu_{\textrm{B}}/\textrm{T}$, for $H_{\textrm{ext}} \parallel [100]$ and $[001]$,
respectively, using the Knight shift values at 10 K\cite{JPSJ.90.023702}.
These values are also evaluated from a macroscopic magnetization\cite{JPSJ.88.054716}.
Note that the present NMR spectra were insensitive to the induced moment
along the [010] direction for $H_{\textrm{ext}} \parallel [100]$, even if it exists.

The uniform and orthogonal $m_{\textrm{ind}}$ at the Ce sites is compatible
with ferroic $O_{zx}$-type quadrupole ordering
among possible quadrupole order parameters summarized in the literature\cite{JPSJ.89.013703}.
In this ordered state, the Ce dipole moment is fixed along the diagonal
direction in the $zx$ plane.
The magnetic field along the [100] ([001]) direction induces the additional
magnetic moment along the [001] ([100]) directions, respectively,
consistent with our observations.
As shown in Fig.~\ref{fig:spectra}(e), no clear anomaly appeared
at $T_{\textrm{s1}}$ in NMR, suggesting that the order parameter emerging below
$T_{\textrm{s1}}$ is insensitive to Co-NMR for ${H_{\textrm{ext}}} \sim [100]$,
which corresponds to $O_{yz}$.
Therefore, it is suggested that phase II' is an $O_{zx}$ ordered state
and phase II is an $O_{yz}+O_{zx}$ ordered state.
The latter gives the triclinic distortion, which is consistent with the XRD
\cite{JPSJ.91.064704}.
This is also consistent with the anisotropic phase diagram between
$H \parallel [100]$ and $H \parallel [110]$ \cite{JPSJ.91.064704}
because $H \parallel [100]$ lifts the degeneracy of $O_{yz}$ and $O_{zx}$.
Additionally, this order parameter completely explains the domain selection
observed in NMR.
When we applied the magnetic field along the [100] direction,
$m_{\textrm{ind}}^c$ was positive in the domains A ($O_{yz}+O_{zx}$)
and B ($-O_{yz}+O_{zx}$),
and negative in the domains C ($O_{yz}-O_{zx}$) and D ($-O_{yz}-O_{zx}$),
as shown in Fig.~\ref{fig:angle-fsp}(a).
Tilting of the magnetic fields towards $\theta_0 < 80$\textdegree\ induces the
positive $c$-axis component in the $H_{\textrm{ext}}$,
stabilizing domains A and B.
In the [110] direction, as shown in Fig.~\ref{fig:angle-fsp}(b),
$m_{\textrm{ind}}^c$ is opposite between the domains A and D.
The domain selection under the $c$-axis magnetic field also occurs when
$H \parallel [001]$.

Ferroic $O_{zx}$ ordering expected in phase II' should induce
a monoclinic distortion.
However, this has not been seen in the $(207)$ reflection of XRD
\cite{JPSJ.91.064704}.
For a setting of the XRD measurement, the order parameter expected from NMR
in phase II' is $O_{yz}$ because the direction of $H_{\textrm{ext}}$
was treated as $[010]$.
In this situation, the domain formation by the monoclinic distortion does not
cause the splitting of the $(207)$ reflection because the mirror symmetry is
not broken with respect to the $yz$ plane.
This might be why the structural change was not seen in the XRD measurement.

The NMR anomalies in the nonmagnetic ordered phase of CeCoSi are totally
comprehended as ferroic quadrupole ordering.
An open issue is the full understanding of the order parameter of this phase.
While the present NMR revealed the existence of the $O_{zx}$ component,
the order parameter could be more complicated than the pure $O_{zx}$ or $O_{yz}+O_{zx}$ state
because of the low crystal symmetry, allowing multiple quadrupole components in the order parameter.
The field-induced successive transitions along the $[100]$ direction could be caused
by the suppression of the $O_{yz}$ component by the field,
although this has not been directly confirmed yet.
A full elucidation of the field--temperature phase diagram is important in CeCoSi.

A more fundamental issue is whether its emergence is possible under the CEF levels
of $\Delta \sim 125$ K \cite{PhysRevB.101.214426}.
Observing the situation under pressure, the difference between $\Delta$ and
$T_0$ becomes smaller ($\Delta \sim 3T_0$)
as $T_0$ increases to 38 K \cite{PhysRevB.88.155137}.
In TmAg$_2$, the quadrupole ordering occurs for $\Delta \sim 3T_0$ even though
the $c$--$f$ hybridization is weak \cite{PhysRevB.48.256}.
Considering the Kondo effect is enhanced under pressure in CeCoSi
\cite{PhysRevB.88.155137,JPSJ.87.023705},
it is not an unrealistic condition for the occurrence of a quadrupole ordering,
although a reason for the high $T_0$ is another problem.
At ambient pressure, weaker $c$--$f$ hybridization suppresses the interaction
between the quadrupole moments and the quadrupole degrees of freedom.
This is consistent with the weak anomalies in bulk properties
at $T_0$ at ambient pressure \cite{JPSJ.88.054716}.
Therefore, the peculiarity of CeCoSi is considered to arise from the
high $T_0$ under pressure.
Ferroic quadrupolar orderings induce lattice distortion as a cooperative
Jahn-Teller effect \cite{PhysRevB.17.3684,J.Phys.Colloques.39.1010}.
Lattice instability, which can assist the ferroic quadrupole ordering,
is an issue to be investigated in CeCoSi.
Note that some tetragonal compounds show lattice distortion to triclinic
symmetry even without a significant contribution from $f$-electrons
\cite{JPSJ.68.2380,APhysPolA.127.219}.
From another perspective, a similarity with CeRh$_2$As$_2$ in the same space
group should also be considered.
These structures do not possess the local space-inversion symmetry of the Ce
site.
Theoretical studies on CeCoSi \cite{JPSJ.89.013703,JPSCP.30.011151} suggested
that the staggered type of an interorbital antisymmetric spin--orbit interaction
stabilized an $O_{x^2-y^2}$-type antiferroic electric quadrupole ordering.
It is an interesting issue to see how these spin--orbit interactions
influence the ferroic quadrupolar interaction.

\section{Conclusion}
In summary, we measured field-angle-controlled $^{59}$Co NMR spectra to reveal
the nature of the nonmagnetic ordered phase in CeCoSi, which possesses
well-separated CEF levels with $\Delta = 125$ K.
The splitting of the NMR peak was interpreted by the formation of domains,
where the Ce dipole moments were induced in different directions
according to the domains.
These features enable the unusual field-driven domain selection of
the nonmagnetic phase.
Our results suggest that the nonmagnetic phase in CeCoSi includes a ferroic
$O_{zx}$-type quadrupole ordering.
The triclinic distortion observed by XRD and no clear anomaly in NMR
at $T_{\textrm{s1}}$ were consistent with $O_{zx} + O_{yz}$-type ordering below
$T_{\textrm{s1}}$.
Our results demonstrate that a quadrupole ordering is possible in a tetragonal system
with large CEF splitting, which is comparable to that of the prototype Ce-based
compound CeCu$_2$Si$_2$.
This suggests that quadrupole interactions can appear in other tetragonal
systems or make a non-negligible contribution behind major magnetic
interactions.

\begin{acknowledgments}
    The authors thank K. Hattori, T. Ishitobi, T. Matsumura, Y. Kawamura,
    M. Yatsushiro, S. Hayami, H. Hidaka, K. Ishida, Y. Kuramoto,
    and K. Fujiwara for their insightful discussions.
    This work was supported by a Grant-in-Aid for Scientific Research on
    Innovative Areas ``J-Physics'' (Grants No.~15H05882, No.~15H05885,
    No.~JP18H04320, and No.~JP18H04321), Grant No.~JP18H03683 and a Grant-in-Aid
    for JSPS Research Fellow (Grant No.~JP19J00336) from JSPS.
\end{acknowledgments}

\section*{Appendix}

\subsection{$^{59}\text{Co}$ nuclear quadrupole resonance and electric field gradient}

Nuclear quadrupole resonance (NQR) measurements were performed on CeCoSi
to obtain information about the structural distortion below $T_0$ under zero field.
The NQR Hamiltonian is
\begin{align}\label{eq:Hamiltonian}
    \mathcal{H}_{\text{Q}} = \frac{h\nu_{\text{Q}}}{6}
    \left[ 3I_z^2 - I(I+1) + \frac{\eta}{2} \left( I_{+}^2 + I_{-}^2 \right) \right],
\end{align}
where $I$ is the nuclear spin ($I=7/2$ for $^{59}$Co), $\nu_{\text{Q}}$ is the quadrupole frequency,
$\eta$ is the asymmetric parameter, and $h$ is the Plank constant.
The Hamiltonian is derived from the electric field gradient (EFG) $V_{ij}$,
which is diagonalized so that $\lvert V_{zz} \rvert \ge \lvert V_{yy} \rvert \ge \lvert V_{xx} \rvert$.
The quadrupole frequency $\nu_{\text{Q}}$ is obtained from the EFG along the
maximum principal axis $V_{zz}$ as follows:
\begin{align}
    \nu_{\text{Q}} = \frac{3}{2I(2I-1)}\frac{eV_{zz}Q}{h},
\end{align}
where $Q$ is the nuclear quadrupole moment and $e$ is the elementary charge.
The asymmetric parameter $\eta$ is defined as
\begin{align}
    \eta \equiv \frac{V_{xx} - V_{yy}}{V_{zz}},
\end{align}
which satisfies $0 \le \eta \le 1$.

The local symmetry at the nucleus is deduced from the NQR frequencies.
The Co site in CeCoSi possesses $\bar{4}m2$ symmetry, ensuring $\eta = 0$ above $T_0$.
In this case, three resonant lines arise at frequencies $\nu_{i} = i\nu_{\text{Q}}$ ($i=1,2$, and 3).
If the four-fold symmetry is broken, $\eta$ emerges, which modifies $\nu_{i}$.
They are given by
\begin{align}
    \nu_1 & = \nu_{\text{Q}}\left( 1 + \frac{109}{30}\eta^2 \right), \\
    \nu_2 & = \nu_{\text{Q}}\left( 2 - \frac{17}{15}\eta^2 \right),  \\
    \nu_3 & = \nu_{\text{Q}}\left( 3 - \frac{3}{10}\eta^2 \right)
\end{align}
within the second-order perturbation with respect to the $\eta$ term in Eq.~\eqref{eq:Hamiltonian}.
The lowest line $\nu_1$ is the most sensitive to the emergence of $\eta$.
Then, the ratio between the first and the second lines is given by
\begin{align}
    \frac{\nu_2}{\nu_1} = 2 - \frac{42}{5}\eta^2.
\end{align}
The value $\nu_2/\nu_1$ is 12 times more sensitive to $\eta^2$ than $\nu_3 / \nu_2$ reported previously \cite{JPSJ.90.023702}.

\begin{figure}
    \centering
    \includegraphics[width=85mm]{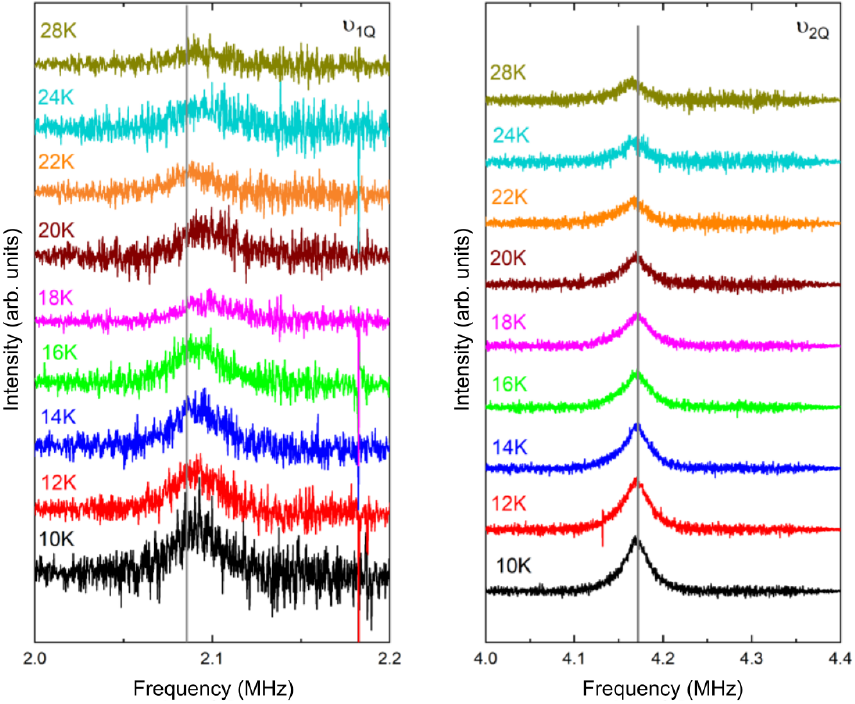}
    \caption{\label{fig:nqrspectrum}
        Temperature dependence of the NQR $\nu_1$ (left) and $\nu_2$ (right) spectra.
        The spectra are shifted vertically.
        The solid lines indicate the case of $\eta=0$ for $T=10$ K.
    }
\end{figure}

\begin{figure}
    \centering
    \includegraphics[width=85mm]{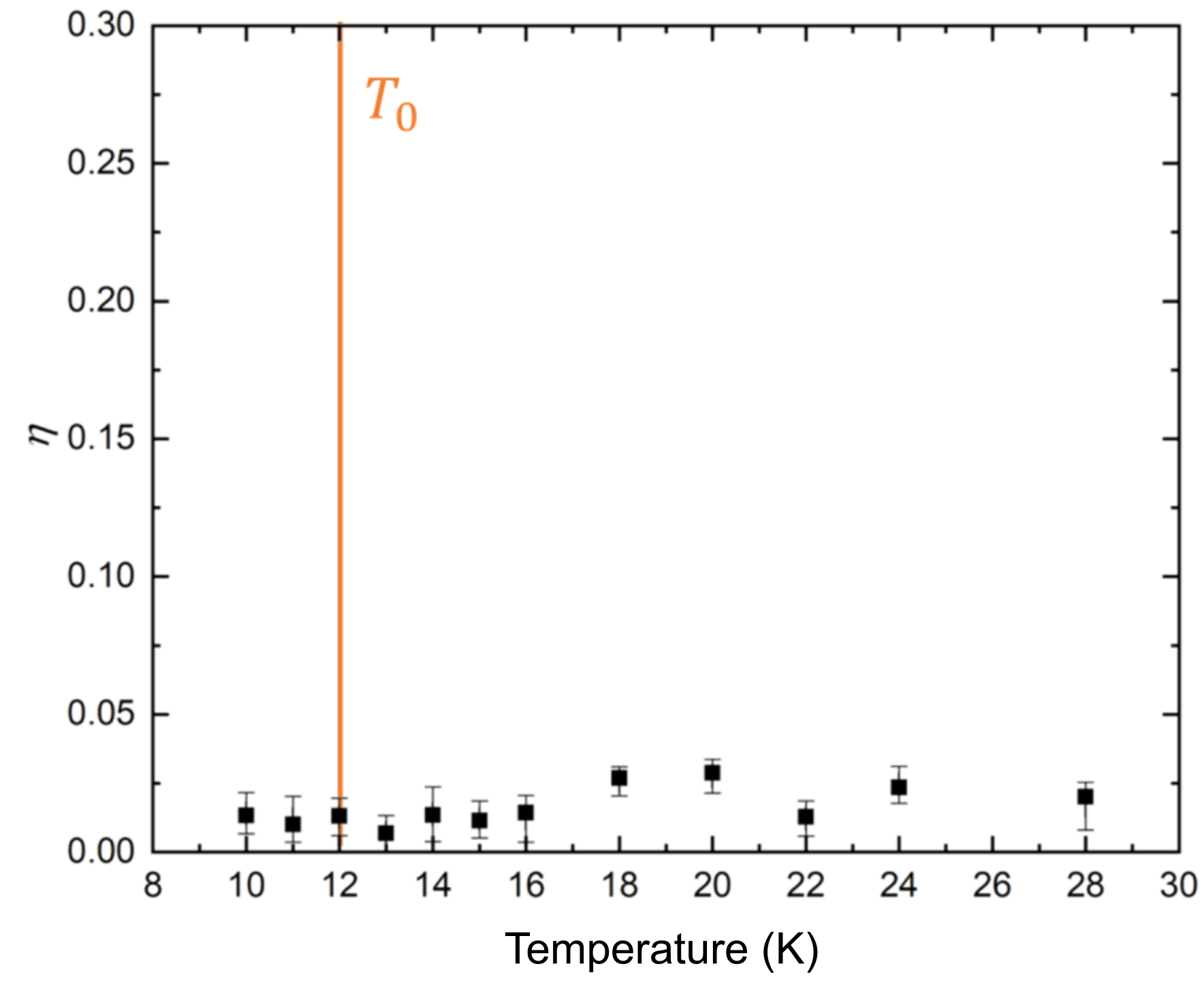}
    \caption{\label{fig:eta}
        Temperature dependence of the asymmetry parameter $\eta$ estimated from
        the NQR $\nu_1$ and $\nu_2$ frequencies.
    }
\end{figure}

Figure \ref{fig:nqrspectrum} shows the NQR spectra arising from $\nu_1$ and $\nu_2$ lines
between 10 and 28 K.
Figure \ref{fig:eta} shows the experimental result $\eta$ estimated from the ratio ${\nu_2}/{\nu_1}$.
No significant increase of $\eta$ was detected at $T_0 = 12$ K, and $\eta < 0.01$ was obtained
as the upper limit.
A small $\eta$ persists above $T_0$.
A possible origin of this is the local inhomogeneity in the $\eta$ term,
which leads to the finite spacial
average of $\eta^2$ even if the spacial average of $\eta$ is zero in the tetragonal phase.
Because the NQR frequency depends on $\eta^2$, the local inhomogeneity could
lead to the residual $\eta$ as a background even in the tetragonal phase.

\begin{table}
    \begin{center}
        \caption{\label{tab:efg}
            The quadrupole frequency $\nu_{\text{Q}}$, asymmetric parameter $\eta$,
            and the angle $\theta_{zz}$ between the maximum principal axis $V_{zz}$
            and the tetragonal $[001]$ axis
            from the band calculation and the NQR experiment at the Co site in CeCoSi.
        }
        \begin{tabular}{cccc}
            \hline\hline
                               & $\nu_{\text{Q}}$ (MHz) & $\eta$  & $\theta_{zz}$ (deg) \\\hline
            Calc. (triclinic at 10 K)  & 2.12                   & 0.013   & 0.16                \\
            Calc. (tetragonal at 20 K)  & 2.11                   & 0   & 0                \\
            Expt. (NQR at 10 K) & 2.085                  & $<0.01$ & ---                 \\\hline\hline
        \end{tabular}
    \end{center}
\end{table}

The EFG parameters are obtained by band calculation through a full-potential linear augmented plane wave
(LAPW) calculation within the local density approximation (LDA).
The lattice parameters are from Ref.~\onlinecite{JPSJ.91.064704}.
The results are summarized in Table~\ref{tab:efg}.
The quadruple frequency $\nu_{\text{Q}}$ was calculated using the quadrupole moment of
$Q = 0.42\times 10^{-28}\ \text{m}^2$ for $^{59}$Co (Ref.~\onlinecite{AtData.111.1}).
The calculated $\nu_{\text{Q}}$ is in good agreement with the experimental one.
The band calculation shows that the asymmetric parameter is $\eta = 0.013$ below $T_0$, which
was in the same order as the experimental upper limit.
The calculation supports the result that the breaking of four fold symmetry was
so small it caused a tiny change in $\eta$.

The tilting of the $V_{zz}$ was also estimated from band calculation.
The obtained value $\theta_{zz} = 0.16$\textdegree\ is much smaller than the experimental NMR split angle
$\Delta \theta \sim 2$--3\textdegree\ at 10 K.
Thus, the NMR anomaly cannot solely be explained by the change in the EFG caused by the structural distortion.

\subsection{Hyperfine coupling between the $\text{Co}$ nucleus and $4f$ electrons}

The hyperfine field at the Co nucleus from the nearest neighbor Ce sites is expressed as follows:
\begin{equation}
    \bm{H}_{\text{hf}} = \sum_{i=1}^{4} \bm{B}_i \cdot \bm{m}_i,
\end{equation}
where $\bm{B}_i$ is the hyperfine coupling tensor between the Co nucleus and the $i$th site,
and $\bm{m}_i$ is the magnetic moment of the $i$th site.
To explain the NMR line split, the following coupling tensor with off-diagonal elements is introduced:
\begin{equation}
    \bm{B}_{1} = \begin{pmatrix}
        B_{aa}^{1} & {\sim 0}   & B_{ac}^{1} \\
        {\sim 0}   & B_{bb}^{1} & {\sim 0}   \\
        B_{ca}^{1} & {\sim 0}   & B_{cc}^{1}
    \end{pmatrix}.
\end{equation}
The coupling tensors for the other sites are obtained by rotating $B_{1}$ with respect to the $c$ axis.
The Ce moments are parallel in the ferroic order, and the hyperfine field is expressed as follows:
\begin{align}
    \label{eq:hyperfine1}\bm{H}_{\text{hf}} & =
    \begin{pmatrix}
        A_{aa}                & {\sim 0}              & B_{ac}^{1}-B_{ac}^{2} \\
        {\sim 0}              & A_{bb}                & A_{ac}^{4}-B_{ac}^{3} \\
        B_{ca}^{1}-B_{ca}^{2} & B_{ca}^{4}-B_{ca}^{3} & A_{cc}
    \end{pmatrix} \cdot \bm{m}, \\
                                            & = A\bm{m},
\end{align}
where
\begin{align}
    A_{aa} & \equiv B_{aa}^{1}+B_{aa}^{2}+B_{bb}^{3}+B_{bb}^{4}, \\
    A_{bb} & \equiv B_{bb}^{1}+B_{bb}^{2}+B_{aa}^{3}+B_{aa}^{4}, \\
    A_{cc} & \equiv B_{cc}^{1}+B_{cc}^{2}+B_{cc}^{3}+B_{cc}^{4}.
\end{align}

The tilting of $H_{\rm total}$ is explained by an emergence of the induced field, ${H_{\text{ind}}}$, at the Co site, whose component is normal to $H_{\textrm{ext}}$.
Since ${H_{\text{ind}}} /{H_{\text{ext}}} \simeq \tan(\Delta\theta)$, the field-independent $\Delta\theta$ indicates that ${H_{\text{ind}}}$ is proportional to ${H_{\text{ext}}}$.
The Ce sites should be uniform even below $T_0$, because the Co sites are uniform.
When ${H_{\text{ext}}}$ is along the $[100]$ axis, the orthogonal ${H_{\text{ind}}}$ at the Co nucleus from the nearest neighbor Ce sites is expressed as follows:
\begin{equation}
    H_{\text{ind}} = A_{cc} m_{c} + A_{ca} m_{a},
\end{equation}
where $m_{c}$ and $m_{a}$ are the $[001]$ and $[100]$ components of the Ce moment, and $A_{cc}$ and $A_{ca}$ are components the hyperfine coupling tensor.
The angle $\Delta\theta \simeq 2$\textdegree\ corresponds to ${H_{\text{ind}}} /{H_{\text{ext}}} \simeq \tan(\Delta\theta) \simeq 3.5$\%, which
is comparable to the in-plane Knight shift $K_{aa} = A_{aa}m_{a}/H_{\text{ext}} \simeq 3.6$\% at 12 K \cite{JPSJ.90.023702}.
It is less likely that the off-diagonal component $A_{ca}$, which was zero in tetragonal symmetry \cite{JPSJ.90.023702}, is comparable to the diagonal part $A_{aa}$.
$H_{\text{ind}}$ is thought to be dominated by the perpendicular moment $A_{cc} m_{\text{ind}}^c$.
Similarly, when the field is along the $[001]$ axis, the induced moment $m_{\text{ind}}^{a}$ causes the induced field
$A_{aa} m_{\text{ind}}^{a}$ at the Co site.

\end{document}